\documentclass[sigconf]{acmart}

\AtBeginDocument{%
  }

\copyrightyear{2023}
\acmYear{2023}
\setcopyright{rightsretained}
\acmConference[CHI '23]{Proceedings of the 2023 CHI Conference on Human Factors in Computing Systems}{April 23--28, 2023}{Hamburg, Germany}
\acmBooktitle{Proceedings of the 2023 CHI Conference on Human Factors in Computing Systems (CHI '23), April 23--28, 2023, Hamburg, Germany}\acmDOI{10.1145/3544548.3581562}
\acmISBN{978-1-4503-9421-5/23/04}

\usepackage{enumitem}
\begin{document}

%%
%% The "title" command has an optional parameter,
%% allowing the author to define a "short title" to be used in page headers.
\title{Using Logs Data to Identify When Software Engineers Experience Flow or Focused Work}

\author{Adam Brown}
\authornote{The first three authors contributed equally to this research.}
\email{adambrovvn@google.com}
\affiliation{%
  \institution{Google}
  \country{USA}}
\author{Sarah D'Angelo}
\authornotemark[1]
\email{sdangelo@google.com}
\affiliation{%
  \institution{Google}
  \country{USA}}
\author{Ben Holtz}
\authornotemark[1]
\email{benholtz@google.com}
\affiliation{%
  \institution{Google}
  \country{Canada}}
\author{Ciera Jaspan}
\email{ciera@google.com}
\affiliation{%
  \institution{Google}
  \country{USA}}
\author{Collin Green}
\email{colling@google.com}
\affiliation{%
  \institution{Google}
  \country{USA}}

%%
%% By default, the full list of authors will be used in the page
%% headers. Often, this list is too long, and will overlap
%% other information printed in the page headers. This command allows
%% the author to define a more concise list
%% of authors' names for this purpose.
\renewcommand{\shortauthors}{Brown, D'Angelo, Holtz et al.}

%%
%% The abstract is a short summary of the work to be presented in the
%% article.
\begin{abstract}
  Beyond self-report data, we lack reliable and non-intrusive methods for identifying flow. However, taking a step back and acknowledging that flow occurs during periods of focus gives us the opportunity to make progress towards measuring flow by isolating focused work. Here, we take
  a mixed-methods approach to design a logs-based metric that leverages machine 
  learning and a comprehensive collection of logs data to identify 
  periods of related actions (indicating focus), and validate this metric against self-reported time in focus or flow using diary data and quarterly 
  survey data. Our results indicate that we can determine 
  when software engineers at a large technology company experience focused work which includes instances of flow. This metric speaks to engineering work, but can be leveraged in other domains to non-disruptively measure when people experience focus. Future research can build upon this work to identify signals associated with other facets of flow.
\end{abstract}

%%
%% The code below is generated by the tool at http://dl.acm.org/ccs.cfm.
%% Please copy and paste the code instead of the example below.
%%


%%
%% Keywords. The author(s) should pick words that accurately describe
%% the work being presented. Separate the keywords with commas.

%%
%% This command processes the author and affiliation and title
%% information and builds the first part of the formatted document.
\maketitle

\section{Introduction}

Flow has been defined as the ``optimal experience'' 
\cite{csikszentmihalyi1990flow} and for decades researchers have 
sought out ways to concretely define, measure, and facilitate  
flow in peoples' lives. An extensive 
literature exists concerning how people experience flow and what factors correlate with these experiences \cite{norsworthy2021review, EFRNscopingreview}. A majority of flow research relies 
on methods employing self-report to understand how and when people experience flow (e.g. 
Experience Sampling Method) \cite{hektner2007experience} which comes with the limitation of only being able to indicate flow after the experience. In an effort to non-disruptively measure flow, researchers have leveraged  physiological data \cite{rissler2018got, maier2019deepflow, rissler2020or}, keystrokes \cite{epp2011identifying} and logs-based data \cite{cowley2022seeking} to measure flow as it is experienced without interruption. However, these studies have had mixed success and the measurement methods--while only mildly intrusive--still require disruption to normal routines (e.g. wearing physiological monitors); disruptions which may adversely affect flow itself. In this work, we sought to measure flow among software engineers (``engineers" throughout the remainder of this paper) through non-intrusive passive behavior tracking via the logs generated from engineering tools. Similar to other professions, engineers stand to benefit from frequent experiences of flow, both in terms of satisfaction with their work and in terms of productivity \cite{murphy2019predicts}. Additionally, while we recognize the ability of survey measures to provide insights into experiences of flow, we sought to develop an "always on" metric that could be informative about experiences of flow without constantly prompting an individual to provide input. Furthermore, having a fine-grained measure for flow enables future research to better understand how targeted interventions can lead to increases in flow. 

We conducted a preliminary diary study in order to understand how flow emerges in the software engineering workflow and to investigate the behavioral signature of flow in logs data. The results of this study drove us to expand the scope of our metric and to consider flow in the context of focused work, with the view that humans achieve flow states if and only if they are doing focused work (i.e., focus is an antecedent to flow), but that they can do focused work without achieving flow (more in Section~\ref{sec:interview_results}). Additionally, and critically for the current work, we found that nearly identical patterns of logs data could be described as "in flow" and "not in flow" based on whether the individual that produced the logs felt positively about the experience in retrospect. These results are not only consistent with the existing flow literature, but also suggest that a logs-based metric for flow (and flow alone) that does not have some input from the individual may not be feasible at this time. Thus, the diary study informed the decision to broaden our scope and develop a metric that captures time spent in focused work, acknowledging that it will capture instances of flow, but that we will be unable to differentiate flow from focused work using only a logs-based approach. We designed and validated this metric against a large-scale diary study and data from a longitudinal survey. The ability to accurately determine when engineers experience focused work gives other researchers the opportunity to combine this metric with other approaches (e.g. self report, physiological data) to better understand when people experience flow through the lens of focused work.

We have defined our metric as \textbf{\textit{focus time}} or time spent engaging in focused work that also includes states of flow. Our preliminary investigation of how engineers experience 
flow indicated a high overlap between what is considered flow and the act 
of engaging in focused work. Consistent with foundational literature on flow \cite{csikszentmihalyi1990flow}, our results suggest that flow is a highly subjective experience, but commonly occurs in the context of focused work (e.g. minimal distractions, ability to concentrate, losing track of time \cite{csikszentmihalyi1990flow}). Therefore, we made the decision to broaden our scope and leverage task relatedness and time spent engaging in related activities to measure when engineers experience focused work. When we discuss our metric we refer to 
time spent in focused work, which may include flow, as \textit{focus time}; we intend for the metric to capture both experiences, but do not attempt to differentiate between the two in this research. Given the eventual goal of identifying a non-intrusive behavioral metric that identifies flow states, we propose that identifying focused work, which at times will detect states of flow, is an appropriate starting point for leveraging a logs-based approach, with future research continuing to identify signatures associated with the other nuanced precursors of flow. 

In this research, we investigate the behavior of engineers at a large 
technology company. Engineers are a suitable population for which to develop a 
logs-based metric for focused work because of the tooling 
used to perform engineering tasks and the logging capabilities of those 
tools. We were well-positioned to conduct this work because of our access to extensive, validated logs data from internal tooling at the company \cite{jaspan2020enabling}. However, the core principles of our metric (time spent engaging in related activities) can be expanded to other types of work and different environments. The ability to measure when engineers experience focused work gets us one step closer to using logs data to identify states of flow and can be used in combination with other approaches to expand our understanding of flow. 

Our multi-phased research addresses the following questions:
\begin{enumerate}
\item How do engineers experience flow?
\item Does the focus time metric reflect the subjective experience of engineers (i.e., does the metric validate against multiple sources of self-reported data)?
\end{enumerate}

Section \ref{prelim} reports a preliminary study that addresses our first research question, while Section \ref{validation} describes two validation analyses for focus time. We then discuss opportunities for future work to better isolate the flow state based on drivers we identified for increasing flow, how flow interacts with productivity, and how the method described can be generalized to other behaviors and contexts. 

\section{Related Work}

Flow and focused work have a long history of research, primarily in the field of psychology. With respect to unpacking and measuring flow and focused work in software engineers and other knowledge workers, Human Computer Interaction (HCI) research has been at the core of understanding how people achieve flow in computer-based work. We discuss the foundational research on flow and focused work, how it has been applied to software engineering, and describe what progress has been made on measuring it. 

\subsection{Flow and Focused Work}

The concept of flow was first articulated by Csikszentmihalyi in the 1970s \cite{csikszentmihalyi1975flowing} and since then has been studied by researchers across a variety of domains. To summarize, flow has been defined as an enjoyable experience engaging in a task that is appropriately challenging and motivating \cite{csikszentmihalyi2005flow, nakamura2014concept, csikszentmihalyi1990flow}. People experience flow across a wide range of activities including hobbies, work, video games \cite{sweetser2005gameflow, klarkowski2015operationalising}, and sports \cite{jackson1999flow}. In the workplace, flow is associated with productivity and satisfaction and is often achieved when a person feels focused and fulfilled with their work \cite{ilies2017flow}. 

We are particularly interested in flow as it occurs in the workplace. Understanding focused work for knowledge workers is critical: multitasking and interruptions are well known to hinder productivity \cite{kim2019understanding} but are common in typical work days \cite{gonzalez2004constant, mark2005no}. In the literature there are many examples of approaches taken to facilitate or encourage focused work. Such approaches include designing interfaces with minimal visual clutter to support focused work \cite{pilzer2020supporting}, blocking or reducing websites unrelated to work \cite{tseng2019overcoming, mark2018effects}, and supporting self-tracking to reflect on time spent and minimize distractions \cite{kim2016timeaware}. Additionally, there has been an increase in the development of time management tools that promote continuous work \cite{manictime_2018, rescuetime._2018}. Together, this research has demonstrated that continuous engagement in related tasks with minimal distractions is important to supporting focused work. This finding--and new research on software engineers, specifically--forms the basis for our design of a metric that hinges on the time spent on related activities.

\subsection{Flow and Focused Work in Engineering}

There is an extensive literature on productivity in software engineering, which considers time spent in flow or focused work to be an important factor \cite{singer2010examination, meyer2014software, murphy2019predicts}. While the current work is not attempting to identify when engineers feel productive, the relevant literature on engineering productivity can help us understand what typical work looks like for software engineers specifically. Especially when taking a logs-based approach, it is important to consider how engineers work. For example, Sanchez et al. \cite{sanchez2015empirical} leveraged Integrated Development Environment (IDE) logs to observe that engineers engage in a high level of activity switching and fragmentation over the course of their day, which had negative impacts on productivity. However, perceived productivity is shown to be very personal and should take into account individual differences \cite{meyer2017work}. Therefore, it is important to understand meaningful changes in activity. Other research has also relied on fine-grained IDE data in order to measure the behaviors of engineers during their workday and make inferences around productive and non-productive behaviors \cite{minelli2015}. While leveraging fine-grained IDE data is useful to \textit{describe} patterns of behavior that have been interpreted as "focus," at no point were the research participants asked to characterize their actual experiences (i.e., we do not know if the patterns described in the IDE data actually felt like focus or flow). 
More recently, Chen et al. \cite{chenworkgraph} developed a measure of focus that leverages logs from a number of work tools (e.g., IDE, chat, internal wiki) by grouping together interactions that do not have interruptions (e.g., a chat message) and present descriptive statistics about how much time engineers spend in focus, as well as what factors impact this time. Similar to the work above, this approach describes focus based on assumptions about which behaviors characterize focus (e.g., using an IDE) and which behaviors do not (e.g., chatting with a colleague). However, we posit that simply using an IDE does not make one focused. The current research aims to integrate fine-grained behavioral data with \textit{reported experiences} of focus and flow, which the research described above has not.

\subsection{Measuring Flow or Focused Work}
Researchers have employed a number of methods to measure when people experience flow, from self-reported survey data and diary studies \cite{hektner2007experience} to tracking physiological data (e.g. heart rate) \cite{rissler2018got, maier2019deepflow, rissler2020or}. However, these approaches inherently disrupt participants' work, sometimes through overt interruptions such as with survey prompts, and sometimes more subtly as with the discomfort or distraction caused by physiological recording devices. In order to study flow uninterrupted and at scale, we sought to employ a non-intrusive and  passive method of data collection; for engineers, logs of activity generated by engineering tools provide a granular and completely passive data collection mechanism, and our approach builds off of this unique opportunity. Recent attempts at defining a scalable logs-based marker of flow using keystroke data concluded that it may not be possible to measure flow accurately from non-intrusively collected data \cite{cowley2022seeking}. Therefore, taking a different lens on how to measure flow by expanding our scope to first measure focus time which includes states of flow appears to be an appropriate first step. Previous research has used physiological data to approach flow from a different perspective by determining when engineers are interruptible or not and avoid disrupting flow states or focused work \cite{zuger2015interruptibility}.  The current work represents a novel approach that builds on existing research and leverages an extensive, granular, and well-validated source of information about engineers' activities derived from logs data to measure flow by measuring focused behavior rather than the subjective experience of flow. These two distinctions from prior work represent the unique contributions of the current research.

\section{Preliminary Study: Understanding How Engineers Experience 
  Flow in Daily Work}
\label{prelim}
Much of the related work has investigated focused work and the subjective experience of flow at a high level across many different tasks and professions. However, in order to define a set of criteria for a metric that indicates when software engineers experience flow in their workday, we first need to understand how engineers experience flow on a more granular level. In this preliminary study, we leverage existing definitions of flow to ask the question, how, if at all, do engineers experience flow or focus in their daily work? 

\subsection{Method}

We recruited 18 software engineers across 6 countries employed at a large technology company to participate in a diary study. Participants were randomly selected from a pool of software engineers who opt in to participating in research and have been at the company for more than 6 months. Prior to participating in the study, participants gave consent allowing collection of their data for the study.

During the five day diary study, participants were instructed to work as they normally would and at the end of each day, to complete an online survey, consisting of the following five questions:

\begin{enumerate}
    \item Did you experience a state of flow while working today? [Yes/No]
\end{enumerate}

\noindent If Yes:
\begin{enumerate}%[resume]
    \item[(2)] How many times did you experience flow?\newline [Once/Twice/More than three times]
    \item[(3)] When did you experience flow? Select all that apply\newline [Morning/Afternoon/Evening]
    \item[(4)] Roughly how long did the flow state last? Select all that apply [Less than 5 minutes/Between 5 - 30 minutes/ Between 30 minutes - 1 hour/Between 1 - 3 hours/More than 4 hours]
    \item[(5)] What activities were you doing while you were in flow? Please be as descriptive as possible.
\end{enumerate}

\noindent If No: 
\begin{enumerate}%[resume]
    \item[(6)] Why didn't you experience flow?
\end{enumerate}

We collected 75 individual diary responses over the week long study period (15 missing diary responses resulted from attrition or non-response from participants). A researcher reviewed the diary responses alongside logs data for each engineer and annotated to indicate when engineers experienced flow and what activities they were doing (details on our logs data can be found in Section~\ref{sec:making sense of logs}). Six participants were randomly selected for a 30-minute semi-structured follow up interview. In these interviews, a researcher and the participant discussed the periods where the participant reported experiencing flow, what logs data indicated they were doing during that time, and ensured we had an accurate interpretation of when the participant experienced flow and when they did not. The interviews were transcribed and paired with diary study responses to determine important characteristics for a logs-based metric.

\subsection{Results}
\label{sec:interview_results}
We conducted a thematic analysis \cite{kiger2020thematic} using inductive coding on the responses to the diary study and follow up interview transcripts to identify common themes that were present across all participants. This resulted in 3 codes for how engineers experience flow, 6 codes for types of tasks participants performed while in flow, and 7 codes for disruptions to flow. One researcher coded all participants data and reviewed decisions with the other researchers. The codes were clustered together to create three primary themes that describe how engineers experience flow:
\begin{itemize}
    \item engineers' experiences of flow are entangled with their judgment of the value of the work or output produced;
    \item engineers experience flow across a range of tasks, not only for specific tasks like writing code;
    \item once established, an engineer's flow can withstand some amount of distraction.
\end{itemize}

\subsubsection{When Engineers Experience Flow}
Our diary study revealed that engineers do not experience flow everyday (70\% of days logged) and when they do experience flow it is typically once per day (62\% of flow days) and in the morning (49\% of flow days). Flow was also experienced across a range of durations with 5 minutes - 30 minutes, 30 minutes - 1 hour, and 1 hour - 3 hours all being similarly common. These diary responses were cross referenced with participant log data and validated in follow up interviews to confirm accuracy. 

When engineers reflected on their flow activities, whether or not they experienced flow was closely tied to how they felt about the activity. As one participant said, \textit{``flow is a visceral feeling.''} Engineers were more likely to express experiencing flow if the outcome was positive. On the other hand, similar activities were cited as not being flow when they experienced frustrations or did not finish the task. 

\begin{quote}
    \textit{``flow didn't come from just coding but making progress on fixing an outage''}
\end{quote}

This is consistent with foundational work on flow \cite{csikszentmihalyi1990flow} that emphasizes the importance of enjoying the activity to experience flow and the subjective nature of flow. However, as our goal is to create a logs-based metric, we took this as a signal that we may not be able to measure flow in its purest form and might need to shift to a broader definition. 

\subsubsection{Tasks That Engineers Consider to be Flow}

We did not identify a consistent set of activities that indicate flow. Our diary study revealed that most engineers experienced flow during coding activities (66\%), but they also expressed experiencing flow during non-coding tasks (e.g. reading documents, planning, or responding to emails). Additionally, engineers often mentioned how they use different but related tools while in flow. For example, in email flow, a participant mentioned seamlessly working through email and docs to accomplish their task while another participant mentioned supporting a coding task by pinging a teammate for help over chat. 

\begin{quote}
    \textit{``I practice `inbox zero', and the process turns email management into a flow-able task.''}
\end{quote}

Engineers also expressed that they were able to maintain flow while experiencing an interruption such as a chat or an email. If these interruptions were easy to address or quick to respond to, engineers were able to remain in flow on their primary task. Interruptions that ended flow were most likely to be from highly unrelated activities (e.g. non-work related interruptions) or longer shifts in focus (e.g. attending a meeting). 

\begin{quote}
    \textit{``I did get interrupted briefly to answer chat, but I was able to quickly resolve it and almost immediately return to a flow state because I had focused on a single task.''}
\end{quote}

Both of these notions appear consistent with the idea of "working spheres" as introduced by Gonzalez and Mark \cite{gonzalez2004constant}. Working spheres represent a set of related events, possibly delineated by tools, that all serve the same motive. For example, in the context of developing a new feature, a working sphere may include the engineer's IDE, existing code repositories, sources of documentation, as well as requirements for the feature that may exist in work tracking software or in a word-processing document. While each of these tools are distinct, they are unified by the task at hand, which allows an engineer to move between them without feelings of fragmentation.

\subsubsection{How Engineers Get Into Flow}

Flow has often been thought of as a state that is achieved through finding your way to the right combination of concentration, skill level, and task difficulty. This puts a larger emphasis on getting the context just right, however engineers in our study expressed having more control over when they achieve flow through goal setting and experiencing flow during tasks that are not typically considered challenging (e.g. email). Throughout the interviews, engineers discussed how they intentionally get into flow. 

\begin{quote}
    \textit{``At the beginning of the cycle, I knew I wanted to implement a fix for a bug to unblock work. I then set my phone to Do not Disturb, played music to my headphones, and began working. I intentionally wanted to get into the flow, and the aforementioned steps helped me reach that state quickly.''}
\end{quote}

Based on the experiences shared, we identified three practices that facilitated flow for engineers:
\begin{itemize}
    \item \textbf{Schedule Management:} Many dispersed meetings prevent engineers from having enough time to get into flow. Establishing a focus or minimal meeting day can help engineers have dedicated time to get into flow. Taking time to reorganize recurring meetings can help prevent dispersed meetings.
    \item \textbf{Goal Setting:} Flow is often achieved when engineers are working on tasks that feel fulfilling. Flow tasks were different across engineers in this study with some experiencing flow solving ambiguous problems while others found flow in more defined tasks. Regardless of task, engineers expressed that having a goal to reach helped get into flow.
    \item \textbf{Time to Get into Flow:} As mentioned, flow can be experienced for varying durations, however regardless of duration, engineers expressed the importance of having sufficient time to “get into flow.” This included setting up their workspace, playing music, and having food/water present. 
\end{itemize}
We discuss opportunities for future research on these practices in Section~\ref{sec:future}. 

In summary, the preliminary diary study provided insights into our first research question: how do engineers experience flow in their daily work? We saw that engineers experience flow across a range of tasks and that flow is robust to small distractions. We also saw that flow is highly subjective, consistent with the prior literature. Based on the results of our study and previous work, we determined that measuring when engineers experience a state of flow using a logs-based approach alone would not be presently possible. However, our study further validates the relationship between experiencing flow and conducting focused work and motivated us to apply our findings to design a metric that gets us closer to measure flow by identifying focused work. 

\section{Focus Time Metric}

\begin{figure}[b]
    \centering
    \includegraphics[width=.3\textwidth]{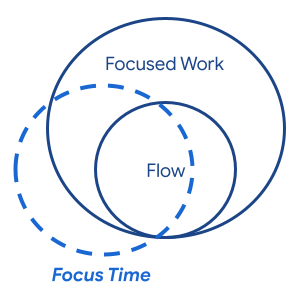}
    \caption{Diagram representing the hypothesized relationship between flow, focus, and focus time}
    \Description{Diagram showing the relationship between flow, focus, and focus time. Focus is a large circle that entirely contains a smaller circle representing flow. Focus time is represented as a third circle that cuts across the intersection of focus and flow, suggesting that sometimes this metric will capture the former, sometimes it will capture the latter, and at times it may capture neither (rote work)}
    \centering
    \label{fig:flocus_circles}
\end{figure}

Based on the results of our preliminary study and the prior literature, we decided to broaden our scope and avoid attempting to differentiate between flow and focused work. Therefore, we sought to develop a metric that (1) is agnostic to task, (2) is robust to small distractions, (3) is flexible on duration, and (4) is independent of the value judgment of the work, rather is reflective of time spent engaging in focused work. We acknowledge that focused work and flow are not the same, but we maintain that being focused on a task is prerequisite to achieving flow. Therefore, identifying when engineers are engaged in focused work can help us capture instances of flow without assigning a value judgment to the task at hand (i.e., we do not need access to an individual's internal states). The remainder of this paper introduces \textbf{\textit{focus time}}: a behavioral measure that indicates when an individual is focused on a task. The results of our preliminary study, paired with the literature on flow, and our goal to develop a passive measure of this phenomenon, has led us to use Figure \ref{fig:flocus_circles} as a conceptual model for how our metric relates to the target constructs that we are interested in. Focus has been repeatedly shown to be a precursor to flow and our preliminary results suggest that in terms of behavioral patterns, focus may generate highly similar patterns of data as a flow state. With this in mind, we consider flow to be a distinct subset of focus. Our behavioral metric, focus time, while relying heavily on the notion of focused behavior, is likely to occasionally detect flow. We also acknowledge that the metric is likely to mislabel behavior that is neither focused nor in flow (e.g., long periods of low activity doing the same thing) or rote work (e.g., common patterns of behavior that occur together that an individual may not find engaging) \cite{mark2014RoteFocus}. Focus time proposes that from a behavioral perspective these behaviors are more similar than they are different, and that leveraging this relationship can help advance research in generating metrics for identifying when individuals experience these constructs. 

The definition of focus time relies on the concept of task similarity (i.e., performing a number of related actions in a given window of time indicates focus time, whereas performing a number of unrelated actions indicates a lack of focus).  We hypothesize that understanding when an individual is focused can also reveal when that individual is experiencing flow. This approach also accounts for our finding that flow is task-independent. In addition it aligns with our finding that flow is robust to small interruptions. 

\subsection{Making Sense of Logs Data}
\label{sec:making sense of logs}
We build on the logs data documented in \cite{jaspan2020enabling}, that are sourced from a diverse suite of development and communication tools used by engineers to complete their day-to-day jobs at a large technology company. These logs are continuously collected to support research questions aimed at understanding engineering behavior at scale. Each log represents a distinct usage of a tool by an engineer or on an engineer's behalf (e.g., running a local test, opening an email, or manipulating version control). Logs are stored, along with relevant metadata, as \textit{events}, standard representations of developer activity. For example, a single event for viewing code within a code search tool will specify the name of the tool, the action associated with the event (e.g. "view"), an identifier for the engineer, the timestamp for when the action started, and metadata that can be useful for further analysis (e.g., the file path associated with the code being viewed). That said, for the purposes of this research, we do not leverage any metadata and rely solely on the tools and actions within those tools to derive our metric. It is also worth mentioning that these logs are subject to a number of privacy principles including, but not limited to, focusing on logs that are associated with work purposes, not collecting user generated content that is not already publicly available, and not reporting out individual data without explicit permission to do so.

In order to define task similarity, we trained a Word2Vec model on events \cite{rehurek2010lrec}. Events were grouped by engineer, ordered by time, and aggregated into sessions broken on 10 minutes of inactivity, similar to previous work using developer logs \cite{jaspan2020enabling}. For example, a session might begin with an engineer making edits within their IDE, then switching over to view some documentation, then returning to their IDE to make additional changes, and eventually trying to build their code. Each of these steps are captured as individual events in our logs and at this step we aggregate them into sessions. Each event gets replaced with a representative label (e.g., "IDE edit") that specifies the name of the tool they are using and the action they are taking. Once constructed, these sessions were treated as sentences of event labels. The model was trained on a 30-day period containing all sessions for all employees within the Software Engineering job profile (i.e., Software Engineer, Site Reliability Engineer, or Research Scientist job codes). This led to $\sim$12 million sessions from $\sim$59 thousand engineers to use as training data for the Word2Vec\footnote{The model was trained using Python's GenSim library \cite{rehurek_lrec}. A skip-gram negative sampling model with a window size of 5 was used to generate 20-dimensional embeddings. Any events that occurred fewer than 200 times were discarded from the training data for the model.}.

The output of this model is a set of custom embeddings for each unique event label that articulates the similarity between the individual activities in our logs data. For example, a search on the intranet is highly similar to reading internal documentation (i.e., these behaviors represent searching for internal information), and both are fairly dissimilar to the command associated with building code. With these embeddings trained, we were able to calculate how similar an engineer's actions are across a span of time, which operationalizes our definition of focus time. 

\subsection{Focus Windows and Focus Sessions}

As mentioned above, our logs data exist as sequences of activities that are generated across a variety of tools by an individual. In order to assign focus values, we first group all events that occur within a sliding window of time (e.g., 10 minutes) and apply a function to generate the similarity between these events in the trained embedding space. 

Formula \ref{eq:focus} represents the \textit{focus time value} ($F$) of a window of events ($W$) as the weighted sum of pairwise distances between all events in the window, divided by the sum of the weights. The weight of events $e$ and $f$ is the product of their respective durations $|e|$ and $|f|$. Event duration is generated by taking the difference between successive event start timestamps. Event distance $dist$ is unit-normalized, so this value is bounded between [0, 1] with 0 representing the perfect focus. In practice, we find better results from also applying a small buffer $B$ to the weights to prevent instantaneous and simultaneous events from being discarded.

\begin{equation}  \label{eq:focus}
    F(W) = \frac{\displaystyle\sum_{e,f \in W}  \left(|e| + B\right)\left(|f| + B\right)dist(e,f)}{\displaystyle\sum_{e,f \in W} \left(|e| + B\right)\left(|f| + B\right)}
\end{equation}

This process assigns a focus value to each event as a representative of the earliest window that includes that event. For example, using the sample session described above, we would take the pairwise similarity between the following events: making an edit within an IDE, viewing documentation, making an edit within an IDE (again), and eventually trying to build the code, and weight this similarity based on the amount of time they spent in these contexts. We hypothesize that these behaviors often occur together as related components of a core development workflow, which translates into these behaviors existing close to one another in the embedding space, thus we anticipate this pattern of behavior to show a focus score closer to 0 (i.e., they are highly related and show less distance between them). We then shift the window to the next set of events and repeat this calculation.

In a subsequent pass, we create \textit{focus sessions} by grouping the focus values of consecutive windows together if their values remain below a specified threshold and they are within a certain amount of time of one another. Focus sessions effectively partition engineers’ days into time in focus and time out of focus. Continuing the example from the previous paragraph, if a developer had a similar set of behaviors in the next window, we would group these windows together into a single session, however, if the next set of behavior had a focus score that was too high (i.e., the behaviors show a lot of distance between them), we would not group them, suggesting that the individual's focus had ended.

Put together, this results in a model with three hyperparameters that we will tune later:
\begin{itemize}
    \item the size of the buffer time $B$,
    \item the minimum duration of the sliding window that events need to be within $W$. 
    \item the threshold for the focus value that events need to stay under for the developer to still be considered in focus 
\end{itemize}

\section{Validating Focus Time}
\label{validation}
While the formulation of the focus time metric was informed by our prior research and the literature, applying our findings to create a formula and generating these values does not--on its own--speak to how well the formulation of the metric actually captures the construct we are interested in measuring. To answer our second research question: does the focus time metric reflect the subjective experience of engineers? We conducted a series of validations with three goals:
\begin{enumerate}
    \item Assess how well focus sessions agreed with whether or not an engineer reported a task as feeling ``focused or in flow'' in a diary study and to identify which combination of hyperparameters led to the highest agreement between these two streams of data;
    \item Rule out naive benchmarks associated with focused behavior (e.g., long sessions of behavior) to show the need for the proposed approach;
    \item Assess whether the optimal version of this metric agrees with self-reported instances of focus and flow collected in a quarterly survey of engineers.
\end{enumerate}

We used two self-reported data sources to validate the focus time metric against different time scales in which participants indicated if they felt in flow or focused. First, we conducted a large-scale diary study, in which participants indicated the time of day and duration for which they experienced flow during one or two work days. Second, we used data from an ongoing quarterly survey in which engineers are asked how often they experienced flow or focus in the past 3 months.    

\subsection{Large-scale Diary Study}

\subsubsection{Method}
We recruited 51 developers across 7 countries employed at a large technology company. Participants were randomly selected from a pool of software engineers who opt in to participating in research and have been at the company for more than 6 months.

\begin{figure*}
    \centering
    \includegraphics[width=.9\textwidth]{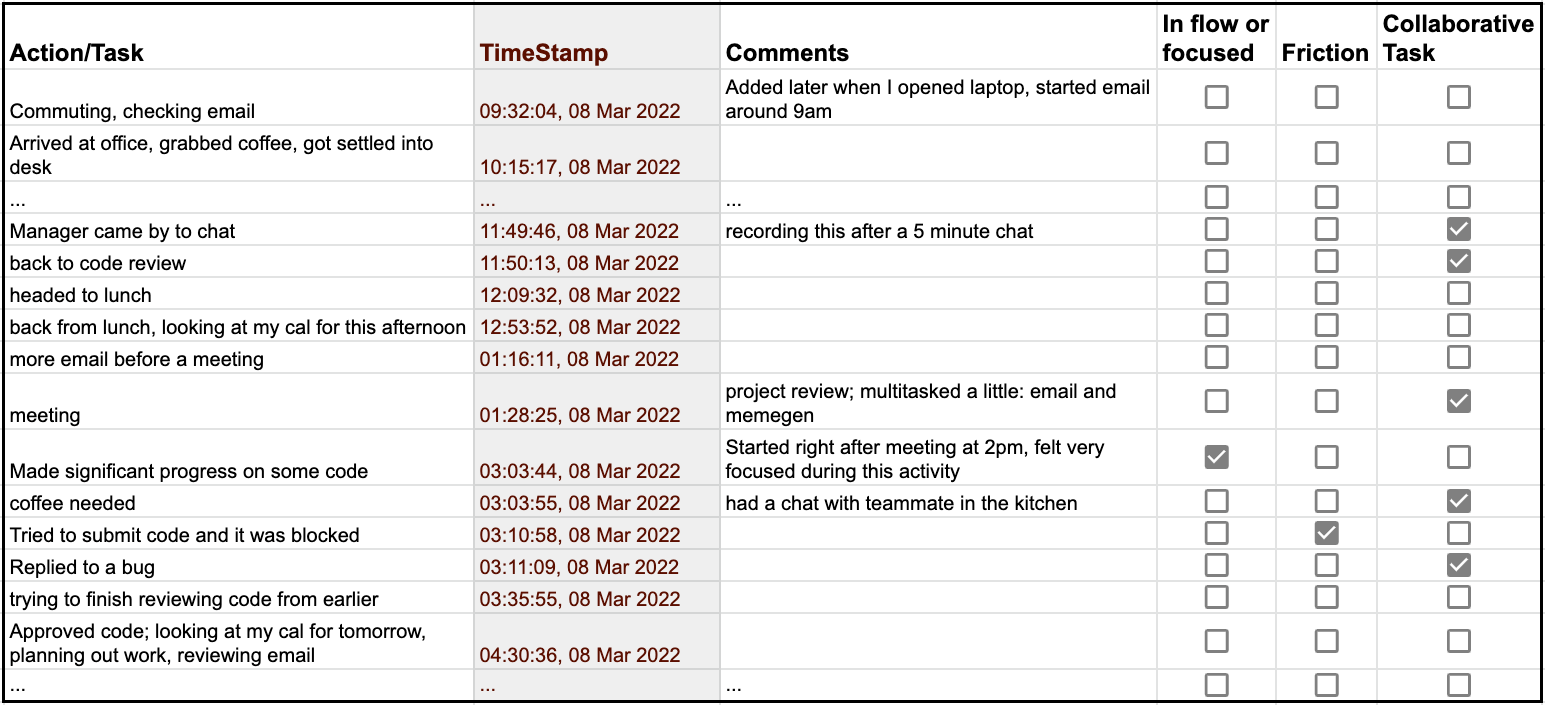}
    \caption{Sample diary study collection form}
    \Description{Sample diary study collection form showing a sheet with activity field, timestamp, comments, and check boxes for in flow/focused, friction, and collaboration.}
    \centering
    \label{fig:sample-diary}
\end{figure*}

During the study, participants were asked to record every activity they did during one or more of their work days using a purpose-built digital form. Forty-six participants recorded two days of work, yielding 97 total days worth of diaries. Participants' diary entries included details on what they did, when they did it, and--critically--whether they felt ``in flow or focused'' during the activity, which they indicated via a checkbox. Figure \ref{fig:sample-diary} shows an example of a diary entry. Participants were instructed to add to their diary entry after every activity they completed during the day. Diaries contained an average of 18 entries per day with an average of 25 minutes elapsed between entries. 

\subsubsection{Validation Analysis}

This data was analyzed using an \newline agreement-based approach by generating whether or not we would label a record in the diary as \textit{focus time} based on how much focus sessions overlapped with the times logged in the diary. The findings presented below will use Prevalence and Bias Adjusted Kappa (PABAK) to measure agreement \cite{byrt1993bias}, which ranges from -1 to +1 with -1 indicating perfect disagreement and +1 indicating perfect agreement. Each diary is given a PABAK value and we will report median PABAK across the entire sample. 

Prior work using this method to analyze diary data for agreement used a benchmarking approach called norm-referencing to interpret PABAK values \cite{jaspan2020enabling}; this method uses relatively straightforward behaviors (e.g. time in meetings or checking email) as the standard for "high agreement" and then compares more complicated behaviors to those standards. For example, checking email showed a median PABAK of 0.84 and time in meetings showed a median PABAK of 0.74 \cite{jaspan2020enabling}, suggesting that behaviors with PABAK near these values are also showing high agreement.

In an attempt to maximize PABAK between the behavioral metric and self-reported focus in the diary data, we conducted a grid search that varied the parameters used to calculate focus sessions and optimized for PABAK. This search manipulated the distance threshold to define focus sessions between [0.1, 0.2, 0.3, 0.4], the length of the focus window between [5 minutes and 60 minutes, using 5 minute increments], and the size of the buffer pad used to give certain instantaneous events some weight between [10 ms and 100 ms, using 10 ms increments]. This led to 480 different tests for agreement to understand how these parameters impacted overall agreement with the diary data. 

Finally, to show the need for the current approach, median PABAK was also calculated for two naive benchmarks that have been hypothesized to be related to the construct of focus time: (1) using sessions of behavior that are longer than the 90\textsuperscript{th} percentile of session length (88 minutes) to define focus and (2) using sessions that have more than the 90\textsuperscript{th} percentile count of events in a session to define focus.

\subsubsection{Results}
\begin{figure}
    \centering
    \includegraphics[width=0.45\textwidth]{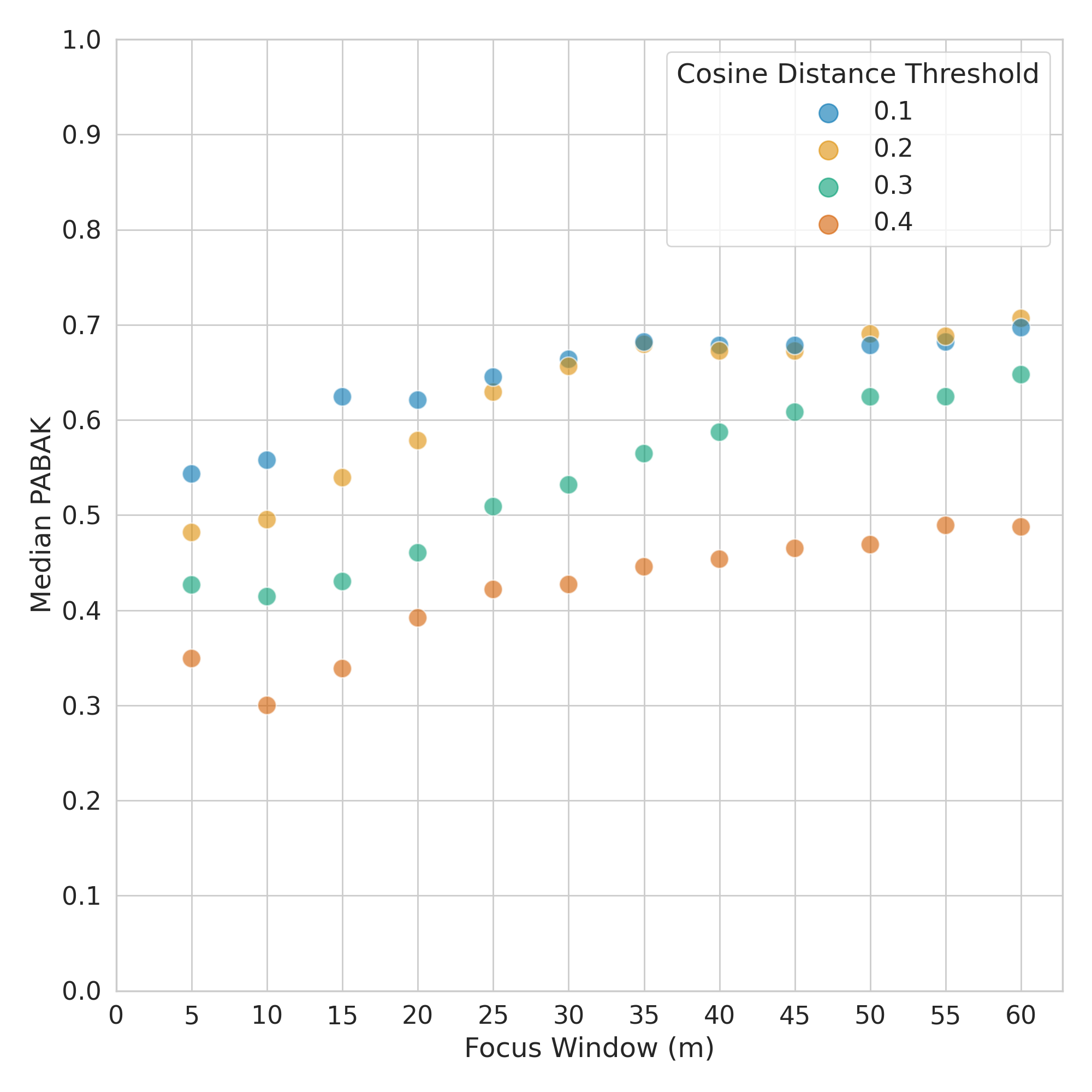}
    \caption{The results of the initial grid search after collapsing across buffer pad values for instantaneous events. These plots show the relationship between window length and median PABAK for each of the distance thresholds used.}
    \Description{A scatter plot showing the results of the initial grid search after collapsing across buffer pad values for instantaneous events. The Y axis shows the median PABAK values from 0 to 1 against the length of the Focus Time Windows from 0 to 60 minutes in increments of 5. There is a series of values shown for each of the distance thresholds used. Lower distance thresholds show consistently higher PABAK values than higher distance thresholds. Longer focus time windows also increase PABAK.}
    \centering
    \label{fig:grid-search}
\end{figure}

\begin{figure*}
    \centering
    \includegraphics[width=.8\textwidth]{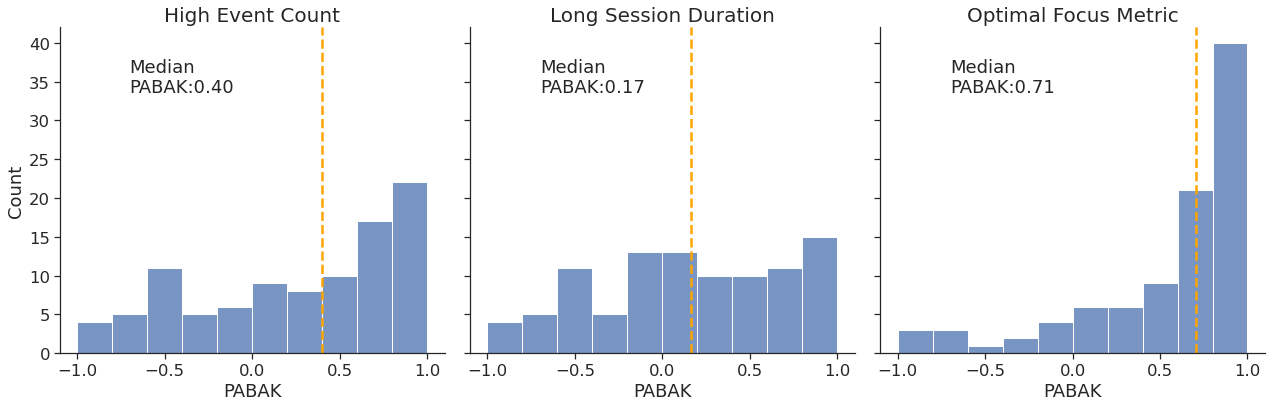}
    \caption{The PABAK distributions for the two naive benchmarks of being in flow or focused and for the optimal focus time metric.}
    \Description{Three bar charts showing the PABAK distributions for the two naive benchmarks of being in flow or focused and for the optimal focus time metric.}
    \centering
    \label{fig:bad-heuristics}
\end{figure*}

The grid search revealed that at the values used, the buffer pad had no impact on median PABAK and was dropped from all remaining analyses; we set the buffer pad to 10ms for the calculation of all focus metrics. The remaining parameters and their median PABAK values are shown in Figure \ref{fig:grid-search}. Generally, it appears that increasing the distance threshold (i.e., allowing for less related activities to be counted as focus) led to lower median PABAK and extending the window of time used increased median PABAK. The maximum median PABAK was 0.71; this version of the metric used a window of 60 minutes and a distance threshold of 0.2.\footnote{Figure~\ref{fig:grid-search} shows diminishing returns in at around a window of 35 min and a distance threshold of 0.2, so we do not feel the need to expand the grid search.} All remaining analyses involving focus time will use these parameters. The PABAK achieved by the optimal focus time metric nearly meets the benchmark set by time in meetings in prior work \cite{jaspan2020enabling}, suggesting that the focus metric generally agrees with engineers' self-report data in the moment.

The two naive benchmarks used for alternative behavioral metrics of focus time showed little-to-no agreement with the diary data. Figure \ref{fig:bad-heuristics} plots the full distributions and highlights the median PABAK values for each benchmark. In both cases, the benchmark used was anticipated to map onto the behavior of focus and is aligned with how individuals discuss the notion focus and/or flow (e.g., ``getting in the zone'' or ``losing track of time''). Sessions that had high event counts had a median PABAK of 0.40, while sessions that were long had a median PABAK of 0.17, suggesting that neither of these adequately capture the behavior of interest. Furthermore, both of these significantly underperform the focus time metric. Taken together, this phase of the validation suggests that the proposed focus time metric maps onto engineers' reports of feeling that they were in flow or focused for a specific task and that the use of custom embeddings is justified based on the poor performance of naive benchmarks. 

\subsection{Quarterly Self-Report Data}

\subsubsection{Method}

In addition to validating focus sessions using fine-grained diary data, we were able to leverage data from a longitudinal survey program that has been conducted at our company for 4+ years. The goals of the survey are to broadly understand engineers' satisfaction and productivity. It contains 50-100 questions depending on branching logic that ranges in topics from tool satisfaction to team communication. Every quarter, the survey is sent out to one-third of eligible engineers (full time employees with a tenure of at least 6 months and a software engineering job code). Respondents are asked to respond to the questions based on their experiences in the 3-month period leading up to the survey. Key to the current research, the survey contains the item: \textit{``How often are you able to reach a high level of focus or achieve `flow' during development tasks?''} Participants answer on a 5-point scale with the options: 1 - Rarely or never, 2 - Sometimes, 3 - About half the time, 4 - Most of the time, or 5 - All or almost all the time. Three consecutive quarters of survey data from Q4 2021 to Q2 2022 were available for this analysis, which functions as a census of the population of interest and contained responses from 13,383 engineers.

\subsubsection{Validation Analysis}
To understand the relationship between focus time and self-reported survey data, focus values need to first be aggregated to the quarter. We generated several ways to aggregate focus time to a quarter:
\begin{itemize}
    \item the total hours spent in focus time in a quarter
    \item the percentage of time spent in focus time (using total session time as a denominator)
    \item the total number of focus sessions in a quarter
    \item the percentage of days in which an engineer had at least one focus session (using active work days as a denominator)
\end{itemize}
Linear regressions\footnote{Linear regression assumes that the distance between flow ratings are equal. Given the question’s wording, we believe this assumption is reasonable.} were run using \texttt{lm} in R to estimate the relationship between quarterly focus time metrics and self-reported focus and flow collected via the survey, while controlling for the following factors: job level, job tenure, job role (i.e., individual contributor, manager, tech lead), and job code (e.g., Software Engineer, Site Reliability Engineer), as well as cohort-level response differences via basis splines.

For the focus time metrics that revealed significant effects, we performed a second set of regressions that included the total number of logged events and the total number of logged sessions an engineer had in a given quarter, in addition to the controls described above. These variables are meant to control for the amount of general activity an individual had in a quarter and help us understand if simply accounting for activity tells us the same information as the focus time metrics. We included the general activity metrics and all control variables in the first regression and added the focus time metric as an additional predictor in the second regression. In these cases, we were particularly interested in whether the focus time metrics significantly predict the self-reported flow item after controlling for this activity, as well as whether the more complex model containing the focus metrics show improved model fit over a  simpler model that does not via F-tests.

\subsubsection{Results}

The results of the linear regression are shown in Table~\ref{tab:stats}.

\begin{table*}
    \centering
    \caption{Linear regression results from the predictive power of four aggregated focus time metrics on focus and flow. For two metrics, this includes controlling for general activity levels.}
    \begin{tabular}{|l|r|r|r|}
       \hline
        Metric & $\beta$ & $t$ & $p$ \\
       \hline
       Hours spent in focus time &  -0.0001 & -1.11 & 0.27\\
        \% of active time in focus time & -0.001 & -1.05 & 0.29\\
        Number of focus sessions & 0.001 & 11.47 & <0.0001 \\
        Number of focus sessions (controlled for activity) & 0.0004 & 8.31 & <0.0001\\ 
        \% of days with a focus session & 0.005 & 6.17 & <0.0001\\
        \% of days with a focus session (controlled for activity) & 0.0004 & 5.21 & <0.0001\\
       \hline
       \end{tabular}
    \Description{Table containing the betas, t-values, and p-values from the quarterly regression analyses.}
    \label{tab:stats}
\end{table*}

\begin{figure*}
    \centering
    \includegraphics[width=.75\textwidth]{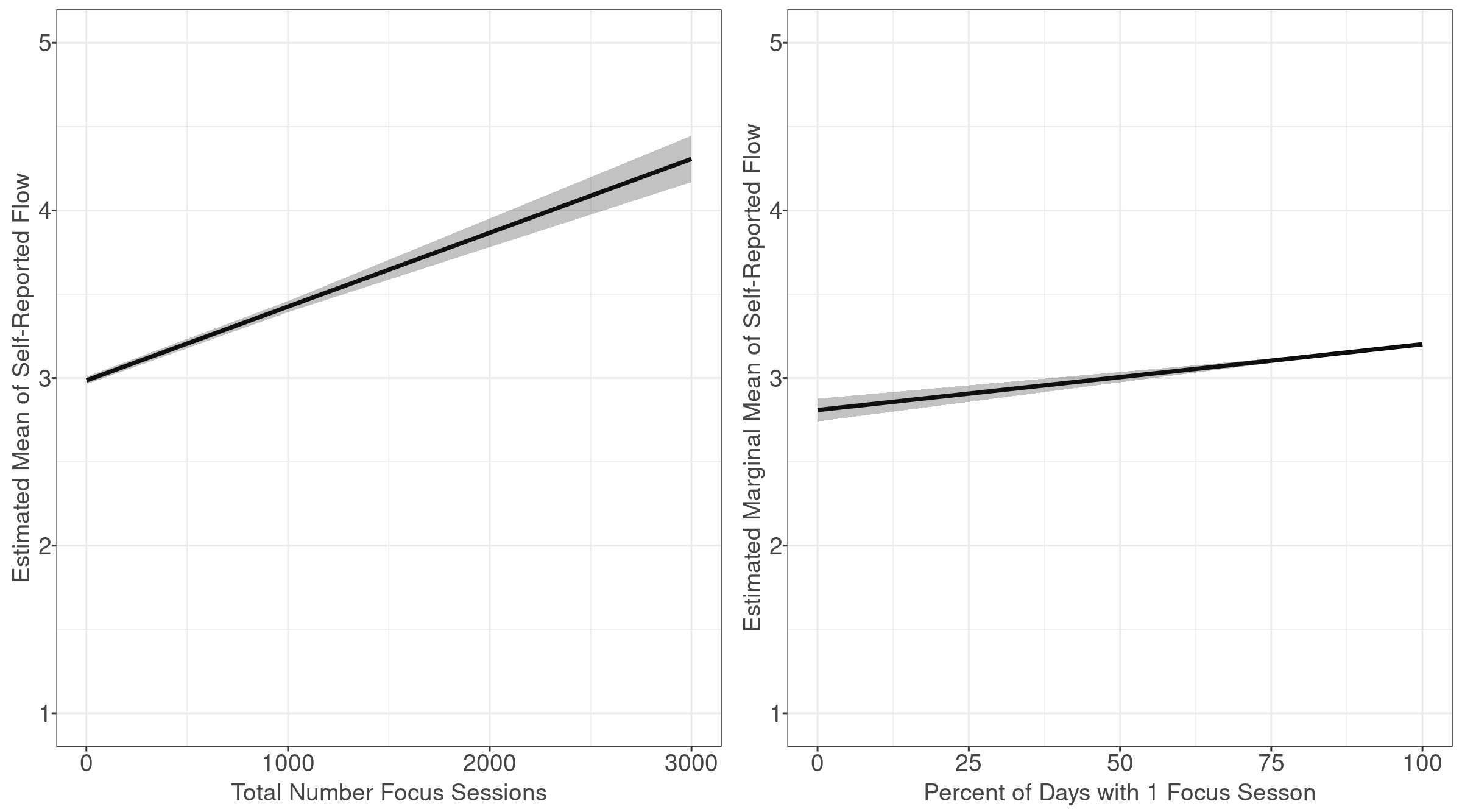}
    \caption{Predicted effects of the focus time metrics on the mean self-reported survey item. These effects were generated at the average values for other continuous predictors and at the reference levels of all other factors, suggesting that these trends are for the typical engineer.}
    \Description{A faceted plot with two line graphs showing the predicted effects of the focus time metrics on the mean self-reported survey item. The Y axis shows the estimated survey score against the focus time metrics on the X axis. Both plots show a positive trend such that increases in the X axis occur with increases in the Y axis.}
    \centering
    \label{fig:engsat-correlation}
\end{figure*}

The total hours spent in focus time in a quarter was not a significant predictor of self-reported focus and flow. Likewise, the percentage of time spent in focus time in a quarter was not a significant predictor of self-reported focus and flow. The implications of the two time-based metrics not showing significant relationships are considered in the discussion section. 

The total number of focus sessions in a quarter was a positive predictor of self-reported focus and flow. This effect remained significant when including the two metrics of general activity described above. Critically, when comparing across the fully specified model and the simpler model that did not include the total number of focus sessions as a predictor, we found statistically better model fit when including the focus metric, $F = 68.99$, $p < .0001$. The left-hand panel of Figure~\ref{fig:engsat-correlation} shows the predicted effect of this focus metric on the self-reported item.

The percentage of days in which an engineer had at least one focus session was also a significant positive predictor of self-reported focus and flow. Again, this effect remained significant when controlling for general activity, and its inclusion in a more complex model led to better model performance, $F = 27.09$, $p < .0001$. The right-hand panel of Figure~\ref{fig:engsat-correlation} shows the predicted effect of this focus metric on the self-reported item.

The results above answer our second research question and support the notion that the focus time metric reflects the experiences of engineers completing focused work and achieving flow.
\section{Discussion}

\subsection{Validating a Behavioral Metric of Flow and Focus}
Across a preliminary diary study, a large-scale diary study, and a longitudinal survey, we were able to identify themes associated with flow and focus in the software engineering workflow, design a metric that maps onto those themes, and validate this metric against self-report data that was recorded relatively close to the end of a given behavior, as well as self-report data that was a reflection on the previous three months.

The results of the large-scale diary data suggest that focus sessions, which were designed to identify periods of time in which an engineer is working on tasks that are largely related, show agreement with periods of time that an engineer reports being in flow or focused. This agreement suggests that at the fairly granular level of diary data, we can accurately identify whether or not an individual will report a given task as feeling "in flow or focused." We also saw that naive benchmarks that are often associated with focus and flow (e.g., longer blocks of activity) showed no to low agreement with diary data, which aligns with previous work aimed at identifying behavioral signatures of flow with similar benchmarks \cite{cowley2022seeking}. Additionally, while other behavioral methods have been able to begin to isolate states of flow \cite{rissler2018got, maier2019deepflow, rissler2020or}, it is important to note that the current approach may represent the first approach to doing so unobtrusively and without requiring further self-report data. Similarly, prior work that has described patterns of fragmentation and focused behaviors have not included any type of quantitative measure for this phenomenon and relied solely on intuitions around what behaviors should and should not count as focus \cite{chenworkgraph, minelli2015, sanchez2015empirical}. The current research introduces a behavioral metric that quantifies focus without requiring these assumptions and instead uses a data-driven approach to identify behaviors that are related, which, as we show through the validation analyses, reveals how focused these behaviors are. 

In a series of analyses using the longitudinal survey responses, we found that quarterly metrics derived from focus sessions were significant predictors of self-reported focus and flow. Most notably, these effects persisted after controlling for general levels of activity, which suggests that simply working more does not fully account for feelings of flow and focus. Furthermore, including focus time variables in these models led to significantly improved model fit. Interestingly, the time-based focus metrics (total time in a quarter spent in focus and percentage of time in focus) were not significant predictors of this outcome. There are at least two possibilities for this outcome that are worth considering. First, the survey question asks about \textit{how often} an engineer is able to achieve a state of flow or focus, so there is some amount of face validity in the focus metrics that center on count or frequency of flow or focus significantly predicting this item, but the duration-based metrics not predicting this item. Second, from the perspective of the literature, flow is associated with subjective distortion of time \cite{csikszentmihalyi1990flow}, which may indicate that there is good reason these two items should not reliably show an association.  

In addition to generating and validating focus time metrics, this research also represents a theoretical shift in how we work towards categorizing and understanding flow and focus behavior in the context of engineering work and--perhaps--more generally. Interviews with engineers during the preliminary study suggested that flow could withstand small interruptions, which appears to be a departure from how flow is typically thought of (i.e., interruptions are often thought of as things that always break flow). The current work also suggests that the similarity of tasks and behaviors may be key elements of understanding flow, or that they function as a good proxy for other factors that typically occur in flow (e.g., having a clear goal may lead to more related actions).

\subsection{Future Research}
\label{sec:future}
\subsubsection{Investigating Factors that Influence Flow}
With the ability to measure focus time, there is a substantial opportunity to investigate what factors drive focus and whether we can increase time spent in focus, which by extension may increase time in flow. Our preliminary study on how engineers experience flow gave us insights into what factors influence how often engineers are able to achieve flow and some practices (schedule management, goal setting) that might facilitate flow for engineers.  We propose investigating the relationship between scheduling dedicated focus time and setting achievable goals on time spent in focus. Recent research has shown that goal setting interventions can help workers experience flow more often \cite{weintraub2021nudging}. We plan to conduct further research to investigate factors such as setting goals and blocking time off on work calendars to understand their impact on focus.

\subsubsection{Developing Additional Signals to Disambiguate Focus Time and Flow}
The current iteration of focus time makes no attempt at separating focus and flow and, instead, aims to leverage the behavioral overlap between these two constructs to provide a general measure of focused behavior that likely detects some instances of flow. However, the other antecedents of flow are nuanced and many definitions of flow include a number of preconditions that expand beyond understanding if an individual is focused (e.g., is the individual adequately challenged?, are there clear goals?, etc.) \cite{EFRNscopingreview, norsworthy2021review}. Focus time represents a first step towards developing a non-intrusive behavioral metric that is capable of handling the nuance of detecting flow. We suggest that additional metrics can (and should) be designed that aim to capture other requirements of flow (e.g., is the current task challenging?) and paired with focus time in order to move research towards a behavioral measure of flow.

\subsubsection{Investigating How Focus Time Relates to Productivity}
Existing research has found associations between \cite{meyer2014software, murphy2019predicts} self-reported feelings of flow and productivity. The focus time metrics allow for future research aimed at understanding whether there are similar relationships between the logs-based metric for this construct and productivity. Future work will assess the relationships between focus and both self-reported productivity and behavioral indicators of productivity typical in software engineering. 

\subsection{Limitations}
The focus time metric is based on extensive logs from internal tools at a large technology company that may not be able to generalize to other companies or types of work. However, this approach represents the first step towards using logs to measure flow and focused work in this way, and researchers can leverage the approach described in this study to evaluate its ability to detect flow and focused work within their domains. Additionally, not all behaviors exist in these logs and there are a number of behaviors that we will never have access to that could influence flow and focus (e.g., interactions with an office environment).

The logs data used in generating our metric, while extensive, lack information about the entity being acted upon. For example, the session used throughout this paper describes an engineer making edits within their IDE, switching over to view some documentation, returning to their IDE to make additional edits, and then building the code. However, it is possible that the first code edit is associated with Project A, the documentation is associated with Project B, and the second code edit is associated with Project B. The current iteration of the metric has no notion of a project and scores focus time based on the overall behavior. In this scenario, we would generate a focus time score that is likely higher than it should be based on there being two potentially unrelated projects involved. However, this limitation is an encouraging one based on our validation results. Given that our current metric already shows strong agreement with engineers' perceptions of focused work and flow, we hypothesize that incorporating this information will only lead to stronger alignment between the metric and the behavior. Additionally, our current results may suggest that this project switching either does not happen frequently enough to reduce agreement values or is not viewed as a large enough change to interrupt an individual's focus.

Furthermore, while providing relatively rich data, the diary study is not without its limitations. First, it may have suffered from a small sample size (\textit{N} = 97). Additionally, one other possible limitation is that bias may have been introduced in the actual tasks that engineers performed on the day they completed the diary. That said, randomly sampling engineers and having them report multiple days, as well as having diaries from different days of the week should have helped reduce this bias. This is one of the reasons why we also performed a validation against longitudinal survey data with a larger sample. 

Another limitation of the current work is that all of our participants had a minimum tenure of 6 months due to sampling practices. It is possible that the current results do not generalize to new engineers or that the results for these individuals may look different than what is reported here. That said, the embeddings that capture task similarity were sampled across all engineers, regardless of tenure, so at a minimum these representations do capture the behavior of new hires. Future work could validate that low-tenure employees show similar patterns of results when compared to high-tenure employees.

\subsection{Generalizing Focus Time to Other Contexts and Data}
Our research presents a new method for using behavioral logs data to generate a metric for capturing flow and focused work. While the current work was able to leverage fine-grained behavioral data from a number of tools that developers use to complete their work, the described approach could be applied to other sources of behavioral logs data that capture human interaction with a technology. One of the potential gaps in previous work that has assessed how focused engineers are based on IDE and tool data is that descriptions of focus are based entirely on mental models and hypotheses rather than quantification of the data \cite{minelli2015, sanchez2015empirical, chenworkgraph}. This method could be applied to these data sources in order to better characterize which behaviors actually reflect focused work and, in the right scenarios, flow. 

However, the method introduced in this research is not limited to data generated from IDEs and other software tools. In fact, any series of behaviors that are ordered by time and associated with an individual could be compatible with this approach. For example, in a more specific context, this method could be applied to clickstream data within a single product or website to generate focus time sessions for that specific interaction. Previous research has used logged online activity to understand times during the day when information workers have more focused attention \cite{mark2014RoteFocus}. Determining the relatedness of websites or activities within a product and leveraging the focus time approach has the potential to improve productivity tools.

There are also a number of opportunities to apply this methodology in other workplaces that have large logs-based datasets to investigate how focused work and flow arise within the specific contexts of these organizations. Once applied, the focus time metric can be used to better understand how much focused work employees experience, as well as the types of behaviors that most often support or interfere with blocks of focused work or flow. Focus time also allows for the exploration of a number of research questions surrounding when focus and flow occur, how long these behaviors last, and what kinds of individual differences impact these experiences. These questions, and likely many more, are of great interest not only to organizations, but to I/O psychologists as well. We propose that our metric can be used to support these investigations.

The focus time metric may also be useful in generating better tools for workers. Specifically, if we are able to measure when an individual is lacking focus or when their focus begins to wane, there is an opportunity to create environments that engage individuals more in an attempt to increase their focus. This metric could potentially support the development of various tools in order to create more engaging content that fosters focus or even be used to investigate what kinds of activities lead to decreases in focus and flow.

\section{Conclusion}
The existing methods available to measure flow or focused work often require explicit prompting or additional monitoring hardware, both of which can be disruptive. A logs-based metric could measure flow and focused work seamlessly and at scale. However, there has yet to be a successful non-intrusive metric that accurately predicts when people experience flow or focused work. In this work, we introduce focus time, a log-based measure of both flow states and focused work. Our results indicate that focus time is accurate, both in the moment as seen in diary data, and across longer periods of time as seen when using data from a quarterly longitudinal survey. We propose that focus time, while intentionally making no attempt to differentiate between flow and focus, represents a meaningful development in the study of flow states. In the short-term, the development of this metric enables researchers to further investigate what drives flow, how to support focused work, and provides a way to measure the success of interventions aimed at increasing flow and focus. In the long-term, using focus time as a baseline measurement, research can be designed to disentangle flow and focus with the goal of developing a behavioral metric that may eventually capture pure flow states. While focus time is currently applied to engineering work, there is also potential to take this approach and apply it to other professions, as well as other types of interactions with technology that capture behavioral logs.  

\begin{acks}
We would like to thank Micheal Bachman, Andrew Macvean, Carolyn Egelman and the Engineering Productivity Research team for their valuable feedback. 
\end{acks}

\balance

%%
%% The next two lines define the bibliography style to be used, and
%% the bibliography file.
\bibliographystyle{ACM-Reference-Format}
\bibliography{flocus-pocus}

%%
%% If your work has an appendix, this is the place to put it.
\appendix

\end{document}